\documentclass[aps,prl,superscriptaddress,twocolumn,longbibliography]{revtex4-2}
\usepackage{xr-hyper}
\usepackage{graphicx, color, graphpap}% Include figure files
\usepackage{enumitem}
\usepackage{amssymb}
\usepackage{amsthm}
\usepackage{amsmath}
\usepackage{dsfont}
\usepackage{mathtools}
\usepackage{soul}
\usepackage[dvipsnames]{xcolor}
\usepackage{multirow}
\usepackage[colorlinks=true,citecolor=blue,linkcolor=magenta]{hyperref}
\usepackage[T1]{fontenc}

\usepackage{thmtools,thm-restate}
\usepackage{verbatim}
\usepackage{titlesec}

\usepackage{graphicx}

\usepackage{bbm}
\usepackage{titlesec}

\usepackage[ruled,vlined]{algorithm2e}

\newcommand{\ket}[1]{|#1\rangle}

\usepackage{etoolbox}
\makeatletter
\patchcmd{\tableofcontents}
  {\relax}
  {\@starttoc{toc}}
  {}{}
\makeatother

\begin{document}

\title{Generation via Classical Noise Reuploading}
% One Step Quantum Generative Model via
\author{Xin Wang}
\affiliation{
	Department of Automation, Tsinghua University, Beijing, 100084, P. R. China
}

\author{Rebing Wu}
\email{rbwu@tsinghua.edu.cn}
\affiliation{
	Department of Automation, Tsinghua University, Beijing, 100084, P. R. China
}

\begin{abstract}
  We propose a novel quantum generative model paradigm that fundamentally avoids the issue of extremely small post-selection probabilities present in previous models. Unlike existing methods that require multi-step noise addition and denoising, this paradigm enables direct single-step generation of quantum data, significantly improving generation efficiency while substantially reducing the complexity of training and quantum state preparation. Furthermore, by directly sampling classical noise to generate quantum states, the sampling process becomes easier to implement. Experimental results demonstrate that this paradigm outperforms existing quantum generative models in terms of generation quality.
\end{abstract}

\maketitle

\noindent\textbf{\textit{Introduction}.--} Quantum generative models inspired by classical diffusion models and flow matching models have recently attracted significant attention. Representative examples include the quantum denoising diffusion probabilistic model (QuDDPM)~\cite{zhang2024generative} and its variants~\cite{kwun2025mixed}, as well as quantum flow matching (QFM)~\cite{cui2025quantum}. These models hold particular promise for generating quantum states that are difficult to prepare: by learning from a small dataset of such states, they can produce a much larger number of samples from the same distribution. Experimental results have demonstrated that these models outperform previous approaches such as quantum generative adversarial networks~\cite{lloyd2018quantum} and quantum direct transport~\cite{zhang2024generative,ding2025quantum} in the task of quantum state generation.

The core idea of QuDDPM is to begin with a training set of quantum states (the target distribution) and transform them into a source distribution that is easy to sample, such as the Haar distribution or the maximally mixed state. This is achieved through a forward process that applies quantum scrambling or depolarizing noise. A backward process, implemented by trainable parameterized quantum circuits (PQCs)~\cite{benedetti2019parameterized} with projective measurements, then learns to generate states whose distribution matches the target, starting from the source distribution~\cite{zhang2024generative,kwun2025mixed}.
QFM addresses two key limitations of QuDDPM: its reliance on both forward and backward processes, and its restriction to special initial distributions (e.g., Haar-random or maximally mixed states), which prevents it from handling simpler sources. QFM instead directly interpolates between the source and target distributions to generate intermediate states, bypassing the forward process. However, because these intermediate states are difficult to prepare and the distance to them is not easily measurable, QFM defines its training loss using a specific metric like entanglement~\cite{cui2025quantum}. In contrast, QuDDPM's forward process makes these intermediate states accessible, allowing its loss function to directly measure distributional fidelity—for instance, via the Wasserstein distance.
To ensure that the generated states faithfully reproduce the target distribution rather than merely optimizing isolated metrics, we adopt the Wasserstein distance as both our primary loss function and evaluation metric.

Due to the inherent linear constraints of quantum channels, it is not possible to directly implement nonlinear transformations between distributions; thus, both QuDDPM and QFM introduce projective measurements to overcome this limitation. The nature of projective measurements means that obtaining the desired quantum state depends on post-selecting certain measurement outcomes, making the quantum state generation process fundamentally probabilistic—in other words, the target state can only be successfully prepared with a certain probability. Moreover, since both QuDDPM and QFM are multi-step generative models and each step involves post-selection, the overall success probability of generating a target state from the source distribution decreases exponentially with the number of steps. In addition, the multi-step generative process increases training costs. These models require not only the preparation of quantum states from the target distribution as a training set, but also additional preparation of intermediate states between the source and target distributions. This runs counter to the original intention of quantum generative models—which aimed to reduce preparation overhead through generation, but in practice end up increasing it for training purposes. Finally, even after the model has been trained, its practical utility remains limited by the difficulty of sampling from the source distribution. Generating quantum states from the target distribution requires the ability to efficiently sample from the source; if the source distribution itself is hard to prepare, the overall practicality of the generative model is greatly diminished.

\begin{figure*}[htpb]
  \centering
  \includegraphics[width=0.95\textwidth]{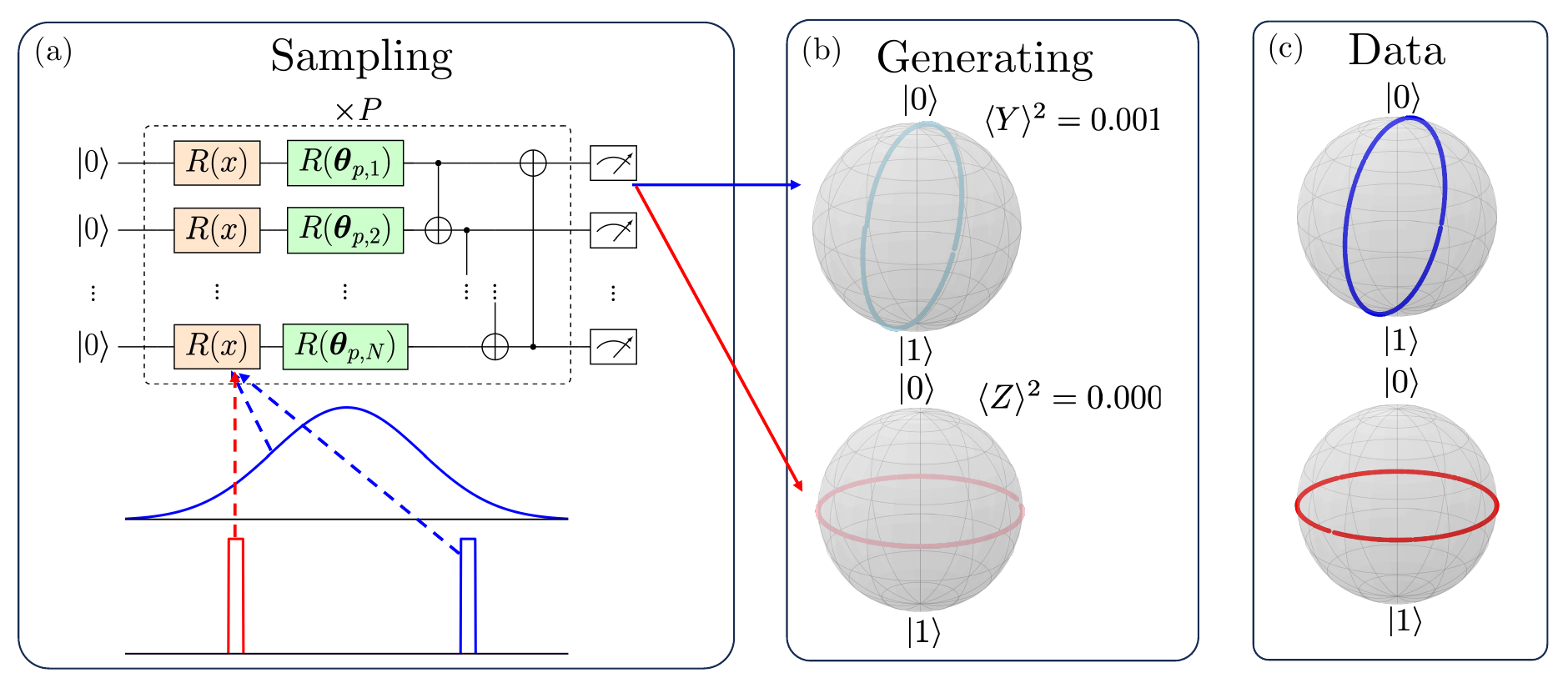}
  \caption{(a) Sampling: Classical noise is sampled from a 1D Gaussian or uniform distribution and inputted into the parameterized quantum circuit. (b) Generating: (single-qubit example) The resulting generated quantum states. The model is capable of generating distinct target distributions---specifically, a ring perpendicular to the Y-axis and another perpendicular to the Z-axis on the Bloch sphere---under the exact same circuit parameters, driven solely by different classical noise samples. Generation quality is evaluated using $\langle Y \rangle^2$ and $\langle Z \rangle^2$, which approach the ideal value of 0 for successful generation. (c) Data: (single-qubit example) The ground-truth quantum state distributions utilized for training the model.}
  \label{fig:illustration}
\end{figure*}
In this work, we propose a novel paradigm for quantum generative modeling. Our model employs a data re-uploading circuit as the generator, where classical noise is used as the input data. By sampling classical noise, the model can generate target quantum states in a single step. We use the Wasserstein distance as the loss function to directly measure and optimize the distance between the re-uploaded quantum states and the target distribution. Our approach offers the following advantages: (1) No post-selection is required, thus avoiding the efficiency issues associated with probabilistic generation; (2) The single-step generation mechanism significantly reduces the training difficulty and greatly improves generation speed; (3) Directly sampling classical noise to generate quantum state distributions greatly enhances the usability and practicality of the model.

\noindent\textbf{\textit{Framework}.--} 
As shown in Fig.~\ref{fig:illustration}(a), we adopt a data re-uploading circuit as the generator. To simplify implementation and avoid potential limitations of data re-uploading models in handling multidimensional independent data~\cite{wang2025predictive}, we set the input noise dimension to one. The circuit takes as input a classical noise variable $x$, which can be easily sampled from a simple distribution such as a Gaussian or uniform distribution. Each gate $R(\boldsymbol{\phi})$ in the circuit denotes an arbitrary single-qubit gate parameterized as:
\begin{equation*}
\begin{aligned}
    R(\boldsymbol{\phi}) = R_z(\phi_{3}) R_y(\phi_{2}) R_z(\phi_{1})
\end{aligned}
\end{equation*}
In the circuit, $\boldsymbol{\theta}_{p,n}$ is a three-dimensional vector representing the trainable parameters for the $n$-th qubit in the $p$-th repetition. For the input $x$, $R(x)$ indicates that all three angles $\phi_1,\phi_2,\phi_3$ in $R(\boldsymbol{\phi})$ are set to this same noise value $x$. The circuit is repeated $P$ times, with the same noise sample $x$ re-uploaded at each repetition, but each time combined with different trainable parameters $\boldsymbol{\theta}$. By optimizing these parameters, the generator learns to map the input noise to the target quantum states.

Following Ref.~\cite{zhang2024generative}, we employ the 2-Wasserstein distance as our loss function. Specifically, given two finite sets of quantum states $S_{1} = \{\ket{\psi_i}\}_{i=1}^{M}$ and $S_{2} = \{\ket{\psi_j}\}_{j=1}^{N}$, we first calculate the inner product $\langle \phi_i|\psi_j \rangle$ between each state in $S_1$ and $S_2$ using the SWAP Test~\cite{barenco1997stabilization}, where $i \in [M], j \in [N]$, and $[M]$ denotes the set $\{1, \cdots, M\}$. We then construct the cost matrix $\mathbf{C}$, with elements $C_{i,j} = 1 - \left| \langle \phi_i|\psi_j \rangle\right|$. The 2-Wasserstein distance $W_{2}(S_{1},S_{2})$ is obtained by solving the following linear programming problem:
$$
\begin{aligned}
  \begin{aligned}
    W_{2}\left(S_{1},S_{2}\right)=\min _{\mathbf{P}} & \langle \mathbf{P}, \mathbf{C}\rangle, \\
    \text{ s.t. } & \mathbf{P} \mathbf{1}_{n}=\boldsymbol{a} \\ 
     &\mathbf{P}^{\top} \mathbf{1}_{m}=\boldsymbol{b} \\
     & \mathbf{P} \geqslant 0 .
    \end{aligned}
\end{aligned}
$$
Here, $\mathbf{P} \geqslant 0$ means that all elements of the matrix $\mathbf{P}$ are non-negative. The vectors $\boldsymbol{a}$ and $\boldsymbol{b}$ are normalized histograms corresponding to $S_1$ and $S_2$, respectively. More specifically, in this work we utilize the Sinkhorn algorithm~\cite{cuturi2013sinkhorn} to enhance both the numerical stability and computational efficiency of the 2-Wasserstein distance.

We further introduce a simple conditional generation mechanism: As shown in Fig.~\ref{fig:illustration}(b), the noise sampling distribution is chosen to be a uniform distribution over two disjoint intervals. For each interval, we assign a corresponding target category of quantum states. With the model parameters $\boldsymbol{\theta}$ fixed, conditional generation can be achieved simply by sampling noise from the appropriate interval according to the desired category and inputting it into the generator circuit. Unlike QFM, which relies on ancillary qubit measurements and post-selection, our approach does not require any additional quantum resources or post-selection operations; instead, different categories of quantum states can be generated solely by explicitly controlling the noise distribution.

\noindent\textbf{\textit{Performance}.--} As a first demonstration, we consider a single-qubit example where quantum states are generated along a ring perpendicular to the $Y$ axis on the Bloch sphere to compare our model with QuDDPM. In the experiment, we prepared a total of 1100 quantum states, using 100 for training and 1000 for evaluation. These quantum states were generated by applying an $R_y(\phi)$ gate to a single qubit, where the angle $\phi$ is uniformly sampled from the interval $[0, 2\pi]$. For classical noise, we chose a one-dimensional noise variable uniformly distributed over $[-0.1, 0.1]$, with the circuit repeated $P = 20$ times. Training was performed using the Adam optimizer with a learning rate of 0.05 for 1000 epochs. During the generation phase, we randomly sampled 1000 times, and the results are shown at the top of Fig.~\ref{fig:illustration}(b). For the entire set of generated quantum states $\mathcal{G} = \{\ket{\psi_i}\}_{i=1}^{M}$, we computed the following metric:
$$
\langle Y \rangle^2 = \frac{1}{M} \sum_{i=1}^{M} \langle Y_i \rangle^2,
$$
where $\langle Y_i \rangle = \langle \psi_i | Y | \psi_i \rangle$ denotes the expectation value of the Pauli $Y$ operator for the $i$-th generated state. Experimental results show that the generated states yield $\langle Y \rangle^2 = 0.001$, which matches very closely with the theoretical ideal value of 0.

\begin{figure}[htpb]
  \centering
  \includegraphics[width=0.45\textwidth]{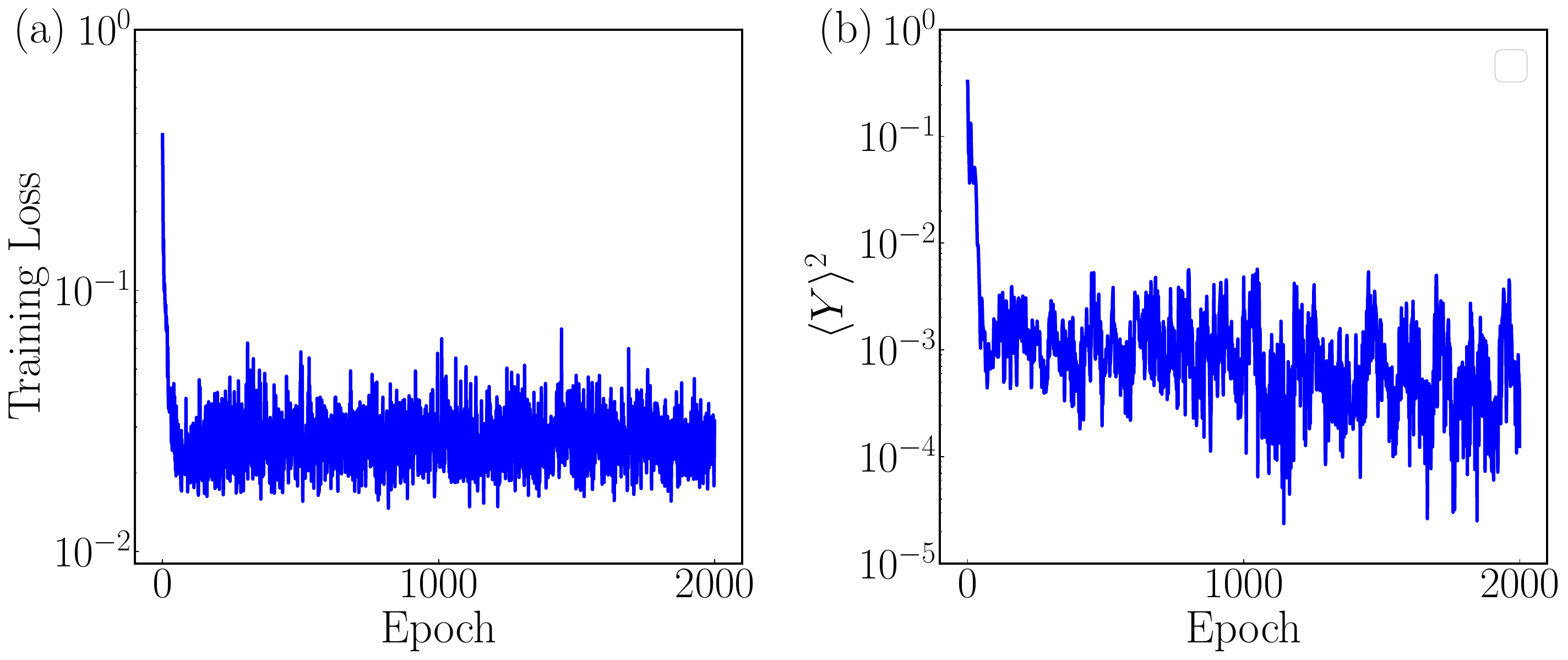}
  \caption{(a) Training loss curve. (b) Evolution of the metric $\langle Y \rangle^2$ as the number of training epochs increases.}
  \label{fig:training_dynamic}
\end{figure}

During training, the evolution of both the training loss and the metric $\langle Y \rangle^2$ of the quantum states in the training set is shown in Fig.~\ref{fig:training_dynamic}. As can be seen from the figure, similar to QuDDPM, the training loss quickly decreases at first and then begins to oscillate.

% Our method naturally enables conditional generation by utilizing different noise intervals. Specifically, we use noise uniformly sampled from the intervals $[-0.1, 0.1]$ and $[0.9, 1.1]$ as inputs. Depending on the interval, the model generates rings on the Bloch sphere that are perpendicular to the Y-axis and Z-axis, respectively. The experimental settings are identical to those for generating the ring perpendicular to the Y-axis alone, and the results are shown in Fig.~\ref{fig:illustration}(b).

Subsequently, under the same experimental settings and using the same training set, we trained the QuDDPM model and compared its generated quantum states to those produced by our model. Unlike the original QuDDPM study, which focused on generalization ability, our work emphasizes evaluating the generative capability of the models. Therefore, we adopt the Wasserstein distance $W_{2}(\mathcal{G},\mathcal{T})$ between the set of generated states $\mathcal{G}$ and the test set $\mathcal{T}$ as the evaluation metric, where both $\mathcal{G}$ and $\mathcal{T}$ each contain 1000 quantum states.

The QuDDPM model is inherently multi-step, and its number of parameters increases with the number of steps $T$. In our experiments, we trained QuDDPM models for $T = 1$ to $20$. In each step, the circuit structure used in QuDDPM is similar to that shown in Fig.~\ref{fig:illustration}(a), except that the data encoding gate $R(x)$ is removed and the circuit depth per step is fixed at 4 layers. To ensure a fair comparison in terms of parameter count, we set the circuit repetition number $P = 4T$ for our model, such that the total number of parameters matches that of the QuDDPM model.

\begin{figure}[htpb]
  \centering
  \includegraphics[width=0.45\textwidth]{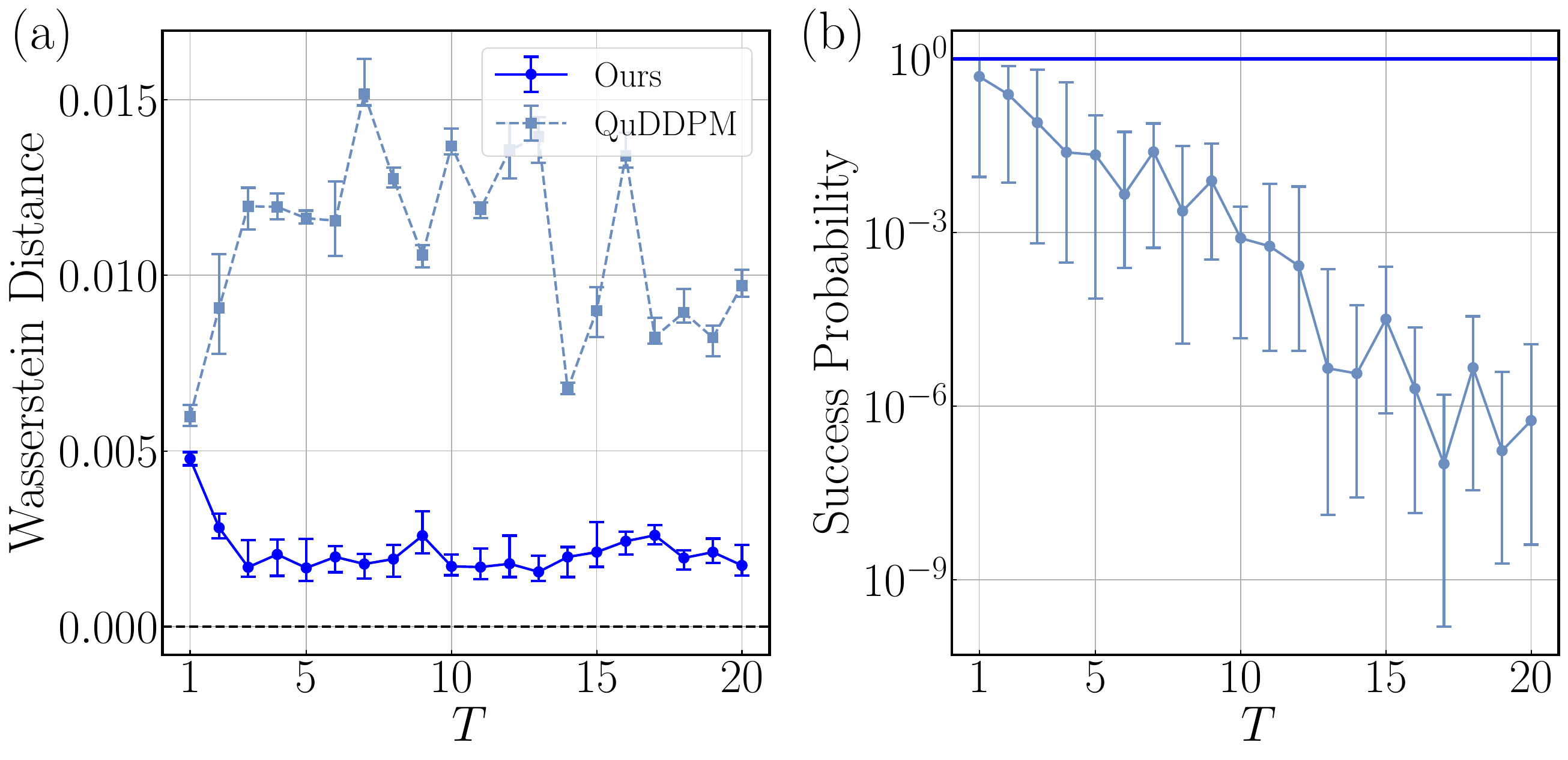}
  \caption{(a) Wasserstein distance between the quantum states generated by our method and QuDDPM and those in the test set as the number of QuDDPM steps (or number of repetitions in our method) increases. (b) Success probability as the number of QuDDPM steps (or number of repetitions in our method) increases. For QuDDPM, the success probability decreases exponentially with the number of steps, whereas for our method it remains 1.}
  \label{fig:comparison}
\end{figure}

The comparison results are shown in Fig.~\ref{fig:comparison}. As can be seen from the results, as the number of repetitions increases, the Wasserstein distance between the quantum states generated by our method and the test set rapidly decreases and then remains stable, and the quality of the generated states consistently surpasses that of QuDDPM. However, as the number of QuDDPM steps increases, its success probability decreases exponentially with the number of steps, while the success probability of our method always remains 1.

\begin{figure}[htpb]
  \centering
  \includegraphics[width=0.45\textwidth]{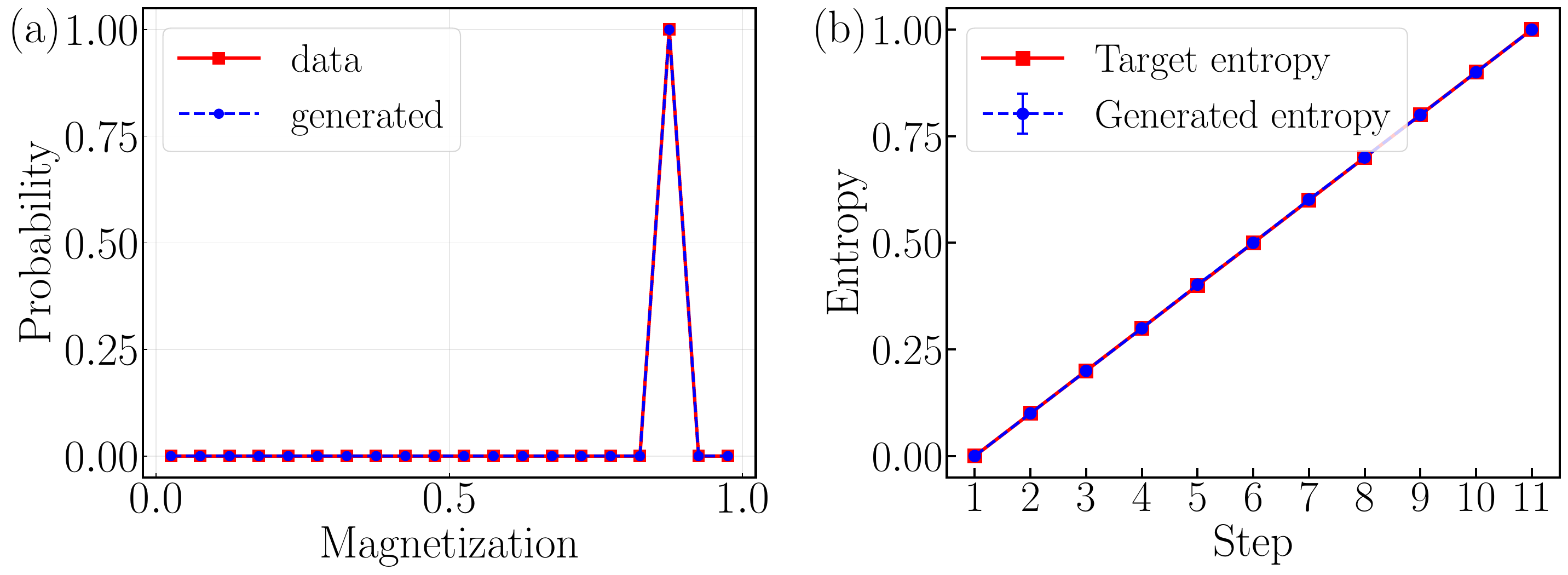}
  \caption{(a) The distribution of magnetization for the generated states matches perfectly with that of the data states. (b) The entanglement entropy perfectly matches the target values.}
  \label{fig:application}
\end{figure}

\noindent\textbf{\textit{Applications}.--} As an example of learning many-body quantum phases, we consider the well-known transverse-field Ising model (TFIM), whose Hamiltonian is given by
$$
H = - \sum_{n=1}^{N-1} Z_n Z_{n+1} - g \sum_{n=1}^{N} X_n,
$$
where $Z_n$ and $X_n$ denote the Pauli Z and Pauli X operators on the $n$-th qubit, respectively. Here, we take the TFIM model with $N=10$ qubits as an example, and choose $g$ uniformly from the range $[1.3, 1.5]$. We randomly sample 100 values of $g$ in this interval and, for each value, obtain the ground state of $H$ to form our training dataset. The 2-Wasserstein distance is adopted as the loss function, with noise uniformly sampled from $[-0.01, 0.01]$. The Adam optimizer is used with a learning rate of 0.05, training for 1000 epochs. During the generation phase, 100 noise samples are uniformly drawn from $[-0.01, 0.01]$ to generate new quantum states. To evaluate the quality of the generated states, we utilize the magnetization $M = \frac{1}{N} \sum_{n=1}^{N} X_n$ to distinguish phases in both generated and data states. As shown in Fig.~\ref{fig:application}(a), the distribution of magnetization for the generated states matches perfectly with that of the data states.

The second task is Entanglement Growth, where we take a two-qubit system as an example to study the evolution of entanglement entropy. The entanglement entropy of the first qubit is defined as
$S(\rho_1) = -\operatorname{Tr}[\rho_1 \log_2 \rho_1],$
where $\rho_1$ is the reduced density matrix of the first qubit.

Let $S_{\text{target}}$ be the target entanglement entropy, and the loss function is defined as
$$
\mathcal{L} = \left( S(\rho_1(\epsilon, \boldsymbol{\theta})) - S_{\text{target}} \right)^2,
$$
where $\rho_1(\epsilon, \boldsymbol{\theta})$ represents the reduced density matrix of the quantum state obtained under noise $\epsilon \in [-0.1, 0.1]$ and circuit parameters $\boldsymbol{\theta}$.

The target entanglement entropy is set uniformly from 0.0 to 1.0 in 11 discrete time steps, with a step size of 0.1. In accordance with the QFM paper, this task does not directly prepare quantum states; instead, it optimizes the circuit parameters to minimize the loss function. The Adam optimizer with a learning rate of 0.05 is used for 1000 epochs. The results from the generation phase are shown in Fig.~\ref{fig:application}(b), demonstrating that the entanglement entropy matches the target values perfectly. Our paradigm achieves this perfect match with straightforward training, without the need for the recurrent switching and selection required by QFM.

\noindent\textbf{\textit{Discussion}.--}  
This paper mainly focuses on the problem of quantum state generation; we do not discuss the transformation of quantum state generation into a classical data generation problem, as there is currently no clear method for decoding quantum states into classical data.

\bibliography{ref}

\end{document}